\documentclass{elsart3p}
\usepackage{natbib}
\usepackage{graphicx}

\newcommand{\dir}{Figs}

\begin{document}

\begin{frontmatter}
\title{A generic model for lipid monolayers, bilayers, and membranes}
\author[A]{Friederike Schmid\thanksref{X}},
\author[A]{Dominik D\"uchs},
\author[A]{Olaf Lenz},
\author[A]{Beate West},
\address[A]{Fakult\"at f\"ur Physik, Universit\"at Bielefeld, Germany}
\thanks[X]{e-mail: schmid@physik.uni-bielefeld.de}

\begin{abstract}
We describe a simple coarse-grained model which is suited to study lipid layers 
and their phase transitions. Lipids are modeled by short semiflexible chains of 
beads with a solvophilic head and a solvophobic tail component. They are forced
to self-assemble into bilayers by a computationally cheap `phantom solvent' 
environment. The model reproduces the most important phases and phase
transitions of monolayers and bilayers. Technical issues such as
Monte Carlo parallelization schemes are briefly discussed.
\end{abstract}
\end{frontmatter}

\section{Introduction}
Lipid bilayers are the main components of biological membranes and omnipresent
in all living matter~\cite{gennis}. At high temperatures bilayers assume the
so-called `liquid' state ($L_{\alpha}$), where lipids are highly mobile and have many 
chain defects. In nature, this state is the most frequent. If one decreases the 
temperature, pure one-component lipid bilayers undergo a prominent phase transition, 
the `main' transition, which is characterized by dropping lipid mobility, dropping 
number of chain defects, and dropping area per lipid. The structure of the low 
temperature `gel' phase depends on the bulkiness and interactions of the head groups. 
For small head groups, the chains are oriented normal to the bilayer ($L_{\beta}$ phase), 
for larger head groups, they are tilted ($L_{\beta'}$). In the latter case, 
the main transition occurs in two steps, and an undulated intermediate phase emerges, 
the `ripple' phase $P_{\beta'}$. If head groups are large and weakly interacting, 
such as ether-linked phosphatidylcholines, the system assumes a phase 
$L_{\beta}^{\mbox{\tiny int}}$ where both opposing lipid layers are fully 
interdigitated~\cite{koynova}. 

In this paper, we present a lipid model which is suited for studying lipid 
bilayers.  We will first apply it to lipid monolayers (Sec.~\ref{sec:lipids})
and show that it reproduces the generic features of fatty acid monolayers. 
Then we introduce an environment model which forces the model lipids to 
self-assemble into bilayers, and discuss the resulting bilayer phases 
(Sec.~\ref{sec:solvent}). Selected technical issues regarding the 
Monte Carlo implementation are discussed in the Appendix.

\section{Lipids and Monolayers}
\label{sec:lipids}

The lipids are represented by chains of $n-1$ `tail' beads with diameter $\sigma_t$, 
attached to one `head' bead with diameter $\sigma_h$. 
Beads that are not direct neighbors along their chain interact with
a truncated and shifted Lennard-Jones potential,
\begin{equation}
\label{eq:vlj}
V_{LJ}(r) =
\epsilon \big[ \: (\frac{\sigma}{r})^{12} - 2 (\frac{\sigma}{r})^6 + v_c \: \big]
\quad \mbox{for} \: r < R_0,
\end{equation}
$V_{LJ}(r) = 0$ otherwise, with $v_c$ chosen such that $V_{LJ}(r)$ is continuous at 
$r=R_c$. The parameter $\sigma$ is the arithmetic mean of the diameters of the 
two interacting beads. Head-head interactions and head-tail interactions 
are purely repulsive, which is ensured by choosing $R_0 = \sigma$. 
Tail-tail interactions have an attractive part, $R_0 = 2 \sigma$. 

\begin{figure}[b]
\vspace{\baselineskip}
\centerline{
\includegraphics[width=0.35\textwidth]{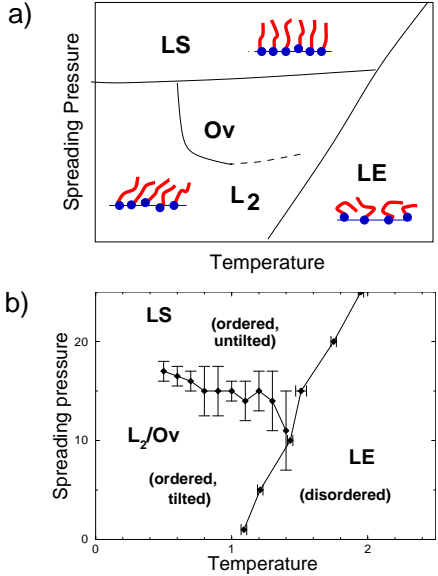}
}
\vspace{-0.5\baselineskip}
 \caption{Monolayer phase diagrams.
(a) Generic phase diagram for fatty acid monolayers 
(after Ref.~\protect\cite{kaganer}). LE is the liquid expanded
phase, the other phases are ordered hexatic liquids. The chains
are untilted in LS, and they tilt in different directions in
L${}_{2}$ and Ov.
(b) Phase diagram of our lipid model. 
From Ref.~\protect\cite{duechs}.
\label{fig:ph_mono}
}
\end{figure}

Within a chain, beads are connected by bonds of length $d$ subject to
the weakly nonlinear spring potential (FENE potential)
\begin{equation}
V_s(d) = 
- \frac{k_s}{2} d_s^2 \ln \Big[ 1\!\! - \!\frac{(d\!\!-\!\!d_0)^2}{d_s^2} \Big] 
\quad \mbox{for} \: |d\!\!-\!\!d_0| < d_s
\end{equation}
and $V_s(d) = \infty$ otherwise, where $d_0$ is the equilibrium spring length, 
$k_s$ the spring constant, and the logarithmic cutoff ensures that the spring 
length never exceeds $d_0+d_s$. 

In addition, a bending potential
\begin{equation}
V_a = k_a (1-\cos\theta)
\end{equation}
is imposed, which acts on the angle $\theta$ between subsequent bonds.

The parameters $\sigma_t$ and $\epsilon$ provide `natural'
length and energy units of the system. In these units, we use
$k_s = 100 \epsilon$ (very stiff bonds), $d_0 = 0.7 \sigma_t$, 
$d_s = 0.2 \sigma_t$, $k_a = 4.7 \epsilon$.
The values are motivated by simple considerations that map our chains
on hydrocarbon chains~\cite{duechs,stadler}; the `matching' should 
not be taken too literally, since the model is not designed to 
describe experiments on a quantitative level. The size of the head group, 
$\sigma_h$, and the chain length, $n$, are model parameters that allow 
to study the influence of the head group bulkiness and the chain length 
on the phase behavior~\cite{stadler}. Unless stated otherwise, they are 
chosen $\sigma_h = 1.1$ and $n=7$.

To evaluate the properties of the lipid model, we first consider monolayers of
lipid at an air-water interface. Such monolayers have been studied for a long time 
as experimentally accessible model systems for lipid layers~\cite{gennis,kaganer}.
The monolayer equivalent of the main transition is a transition encountered
upon compression of the monolayer from a `liquid expanded' (LE) phase to a 
`liquid condensed' phase. As in bilayers, the ordered `liquid condensed' phase exists 
in several modifications, which differ, among other things, in the tilt order
of the chains (L${}_2$, LS, or Ov phase, see Fig.~\ref{fig:ph_mono}a).
In the simulations, the water surface can be replaced by suitable external potentials. 
With smooth harmonic potentials of width $\sim \sigma$, we obtain the phase diagram 
shown in Fig.~\ref{fig:ph_mono}b)~\cite{duechs}. It is in good qualitative agreement 
with the experimental phase diagram, Fig.~\ref{fig:ph_mono}a).

\section{Phantom solvent and self-assembly}
\label{sec:solvent}

\begin{figure}[tb]
\centerline{
\includegraphics[width=0.35\textwidth]{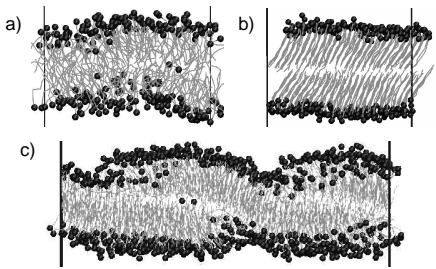}
}
 \caption{Snapshot of lipid bilayers 
  a) fluid bilayer $L_\alpha$,
  b) tilted gel $L_{\beta'}$
  c) asymmetric ripple $P_{\beta'}$
\label{fig:snapshots}
}
\end{figure}

Having formulated a reasonable lipid model, we must now force the
`lipids' to self-assemble into bilayers. In nature, self-assembly
is caused by the interaction with the surrounding water, hence we
must add an appropriate representation for the environment. This
is done by introducing a recently proposed, simple and very efficient 
environment model: The `phantom solvent' model~\cite{lenz}. Explicit
`solvent' particles are added to the system, which however do not
interact with each other, only with lipid beads (by means of repulsive 
interactions, Eq.~(\ref{eq:vlj})). Physically, the solvent 
probes the accessible free volume in the presence of lipids on the length 
scale of the solvent diameter $\sigma_s$. Therefore, it promotes lipid aggregation, 
and the lipids self-assemble to bilayers (see Fig.~\ref{fig:snapshots}). 
Compared to other explicit solvent models~\cite{explicit}, the phantom 
solvent environment has the advantage of having no internal structure, 
it thus transmits no indirect interactions between different bilayer 
regions and/or periodic images of bilayers. Furthermore, it is cheap 
-- in Monte Carlo simulations, less than 10 \% of the computing time is 
spent on the solvent. Compared to implicit solvent models~\cite{implicit},
where the solvent is replaced by effective lipid interactions,
it has the advantage that no tuning of potentials is required,
and it can also be used to study solvent dynamics.
For example, with DPD dynamics, one can study the effect of
hydrodynamic coupling between membranes and the surrounding fluid.

In our work, the solvent diameter was chosen $\sigma_s = \sigma_h$. Single head 
beads are soluble, ({\em i.e.}, they do not demix with solvent) if the free 
solvent density is less than $\rho_{\mbox{\tiny free}} \sim 2.6/\sigma^3$. 
At sufficiently low temperatures, the lipids self-assemble into bilayers 
(see Fig.~\ref{fig:snapshots}). The properties of these bilayers will be
discussed in detail elsewhere~\cite{lenz2,lenz3}. Here we just cite
some of the main results. Like the monolayers, the bilayers
exhibit a main transition. For small heads $(\sigma_h = 0.9\sigma)$, 
the gel phase is untilted, {\em i.e.}, we obtain an $L_{\alpha}$ phase. 
For larger heads $(\sigma_h = 1.1 \sigma)$, the structure of the gel 
phase depends on the free solvent density $\rho_{\mbox{\tiny free}}$.
We note that the solvent entropically penalizes lipid/solvent 
interfaces and thus effectively creates an attractive depletion 
interaction between the beads next to these interface, {\em i.e.}, 
the head beads. The strength of this interaction is proportional 
to $\rho_{\mbox{\tiny free}}$. At low $\rho_{\mbox{\tiny free}}$, \
the gel phase is interdigitated ($L_{\beta}^{\mbox{\tiny int}}$),
at moderate $\rho_{\mbox{\tiny free}} > 1.2/\sigma^3$, it is tilted 
($L_{\beta'}$). Hence weak head attraction leads to the formation 
of the interdigitated phase, and moderate head attraction to the 
tilted gel phase, in agreement with experiments.

Most rewardingly, we also recover the ripple phase $P_{\beta'}$
which intrudes between the tilted gel phase and the fluid phase.
A snapshot is shown in Fig.~\ref{fig:snapshots} c). 
The two main experimental rippled states, the `asymmetric' and the 
`symmetric' rippled state, are recovered in simulations, with properties 
that are very similar to experimental properties~\cite{lenz2}. 
A similar structure than that of our asymmetric ripple has been
found recently in a (much more involved) atomistic simulation
of a Lecithine bilayer~\cite{devries}. Our simulations show that this 
structure is generic, in the sense that it can be reproduced with a 
coarse-grained model, and that it is closely related to the
structure of the corresponding symmetric rippled state~\cite{lenz2}.

\section{Conclusions and Outlook}
\label{sec:conclusions}

\begin{figure}[tb]
\centerline{
\includegraphics[width=0.35\textwidth]{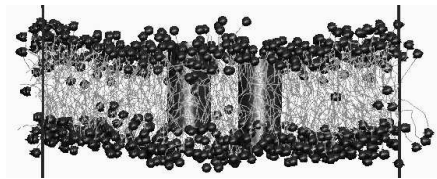}
}
 \caption{Fluid membrane with two embedded coarse-grained
transmembrane proteins
\label{fig:protein}
}
\end{figure}

To conclude, we have presented a versatile coarse-grained 
model that allows to study lipid monolayers and self-assembled
bilayers and reproduces their most important internal phase transitions. 
It can be used to study a variety of questions related to
membrane biophysics where atomic details do not matter, 
but the characteristic molecular features of lipids are 
still important. For example, we are currently applying it
to study lipid-mediated interaction mechanisms between proteins
An example snapshot is shown in Fig.~\ref{fig:protein}.

\bigskip

\noindent
{\bf Acknowledgements} 
\bigskip

We thank the NIC computing center in J\"ulich for
computer time. This work was funded by the Deutsche
Forschungsgemeinschaft.

\bigskip
\noindent
{\bf Appendix: Technical Remarks} 
\bigskip

The Monte Carlo simulations described above were carried out at constant 
pressure $P$ with periodic boundary conditions in a simulation box of variable 
size and shape: The simulation box is a parallelepiped spanned by 
the vectors $(L_x,0,0)$, $(s_1 L_x,L_y,0)$, and $(s_2 L_x,s_3 L_y, L_z)$.
All $L_\alpha$ and $s_i$ are allowed to fluctuate. In addition, it is sometimes 
convenient to work in a semi-grand canonical ensemble with fluctuating number 
$N_s$ of solvent beads and given solvent chemical potential $\mu_s$. Hence we have 
three types of possible trial Monte Carlo moves:
\begin{itemize}
\item Moves that change the positions of beads.
\item Moves that changes the volume and/or shape of the box: Random increments
drawn randomly from a symmetric distribution with mean zero are added to 
$L_\alpha$ or $s_i$. All bead coordinates are rescaled accordingly.
\item Moves that changes the number $N_s$ of solvent particles: We first
decide with probability 1/2 whether to attempt a solvent removal or a 
solvent addition. Then, we choose randomly the particle to be removed,
or the position of the particle to be added.
\end{itemize}
The moves are accepted or rejected according to one of the standard 
Monte Carlo schemes ({\em e.g.}, Metropolis), with the effective 
Hamiltonian~\cite{frenkel_book}  
\begin{equation}
H_{\mbox{\tiny eff}} = H \!+ \!P V \! -\! \mu_s N_s
\!-\! k_B T \ln \big[(V/V_0)^N/N_s!\big],
\end{equation}
where $H$ is the interaction energy, $V = L_x L_y L_z$ the volume of the
simulation box, $V_0$ an arbitrary reference volume ({\em e.g.}, $V_0 =
\sigma_t^3$), and $N$ the {\em total} number of beads (solvent and lipids). 
The $L_\alpha$ are not allowed to fall below a given threshold, otherwise
the move is rejected.

\begin{figure}[tb]
\centerline{
\includegraphics[width=0.25\textwidth]{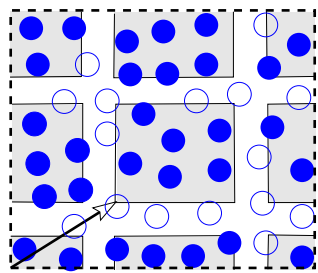}
}
 \caption{Sketch of the domain decomposition scheme used
for the parallelization of the Monte Carlo algorithm
(a variant of Uhlherr \etal~\protect\cite{uhlherr}).
The system is covered with a grid of active regions 
(shaded rectangles, filled particles). 
The distance between active regions must exceed the 
interaction range between beads. The offset of the 
active grid (arrow) changes periodically and is chosen 
randomly.
\label{fig:parallel}
}
\end{figure}

We close with a remark on parallelization. For large scale applications, 
our Monte Carlo code has been parallelized geometrically.
One commonly used spatial decomposition scheme for systems with short range 
interactions (see, {\em e.g.}, the review~\cite{heffelfinger}) proceeds as follows: 
The simulation box is divided into domains 
which are distributed on the processors. These are further subdivided into 
labelled subdomains such that subdomains with the same label are separated 
by a distance larger than the maximum interaction range. Subdomains 
with the same label are then processed in parallel. This algorithm is 
relatively straightforward, yet it has the drawback that it does not 
strictly fulfill detailed balance: Within a move for given subdomain 
label $\alpha$, particles can cross a subdomain boundary only in one 
direction ({\em i.e.}, leaving the set of subdomains $\alpha$). If
the different sets $\alpha$ are processed equally often, the final distribution 
is presumably not affected. Nevertheless, we feel uncomfortable with
this method and favor a variant of a parallelization scheme recently proposed 
by Uhlherr \etal~\cite{uhlherr}. The idea is to define `active regions' 
and assign them to different processors.  A possible decomposition scheme is 
shown in Fig.~\ref{fig:parallel}. Only particles with centers in the active 
regions are moved, and moves that take a particle outside of its active 
region are rejected. To ensure that the algorithm remains ergodic,
the active regions are periodically redefined. One easily checks
that individual bead moves fulfill detailed balance.
We note that the active regions must not necessarily have the same size, 
and furthermore moves in interesting regions can be more frequent. 
This feature makes the concept of `active regions' interesting even
for applications on scalar computers.

\noindent
{\bf References}

\vspace*{-3\baselineskip}


\begin{thebibliography}{9}

\bibitem{gennis}
  R.~B. Gennis, {\em Biomembranes},
  Springer Verlag, New York, 1989.

\bibitem{koynova}
 R. Koynova, M. Caffrey, {\em Chem. Phys. Lipids} {\bf 69}, 1 (1994);
 {\em Biophys. Biochim. Acta} {\bf 1376}, 91 (1998).

\bibitem{kaganer}
 V. M. Kaganer, H. M\"ohwald, P. Dutta, {\em Rev. Mod. Phys.} {\bf 71}, 779
 (1999).

\bibitem{duechs}
  D. D\"uchs, F. Schmid, {\em J. Phys.: Cond. Matter} {\bf 13}, 4835 (2001).

\bibitem{stadler}
  C. Stadler, H. Lange, F. Schmid, {\em Phys. Rev. E} {\bf 59}, 4248 (1999);
  C. Stadler, F. Schmid, {\em J. Chem. Phys.} {\bf 110}, 9697 (1999).

\bibitem{lenz}
  O. Lenz, F. Schmid, {\em J. Mol. Liquids} {\bf 117}, 147 (2004).

\bibitem{explicit}
 B. Smit \etal, {\em J. Phys. Chem.} {\bf 94}, 6933 (1990);
 R. Goetz, R. Lipowsky, {\em J. Chem. Phys.} {\bf 108}, 7397 (1998);
 J.~C. Shillcock, R. Lipowsky, {\em J. Chem. Phys.} {\bf 117}, 5048 (2002);
 M. Kranenburg, J.~P. Nicolas, B. Smit, {\em Phys. Chem. Cehm. Phys.} {\bf 6},
 4142 (2004).

\bibitem{implicit}
 H. Noguchi, M. Takasu, {\em Phys. Rev. E} {\bf 64}, 041913 (2001);
 O. Farago, {\em J. Chem. Phys.} {\bf 119}, 596 (2004);
 I.~R. Cooke, K. Kremer, M. Deserno, {\em Phys. Rev. E} {\bf 72}, 011506 (2005).

\bibitem{lenz2}
 O. Lenz, F. Schmid, submitted (2006) {\tt www.arxiv.org/abs/physics/0608146}.

\bibitem{lenz3}
 O. Lenz, F. Schmid, in preparation.

\bibitem{devries}
 A.~H. de Vries \etal, {\em PNAS} {\bf 102}, 5392 (1005).

\bibitem{frenkel_book}
  D. Frenkel, B. Smit, {\em Understanding Molecular Simulations},
  Academic Press, San Diego, 2002.

\bibitem{heffelfinger}
  G.~S. Heffelfinger, {\em Comp. Phys. Comm.} {\bf 128}, 219 (2000).

\bibitem{uhlherr}
A. Uhlherr \etal, {\em Comp. Phys. Comm.} {\bf 144}, 1 (2002).

\end{thebibliography}
\end{document}